# Frustrated magnetic interactions in a Wigner-Mott insulator

Yanhao Tang[1,2], Kaixiang Su[3], Lizhong Li[1], Yang Xu[1], Song Liu[4], Kenji Watanabe[5], Takashi Taniguchi[5], James Hone[4], Chao-Ming Jian[6], Cenke Xu[3], Kin Fai Mak[1,6,7], Jie Shan[1,6,7]

[1]School of Applied and Engineering Physics, Cornell University, Ithaca, NY, USA
[2]Interdisciplinary Center for Quantum Information, Zhejiang Province Key Laboratory of Quantum Technology, and Department of Physics, Zhejiang University, Hangzhou, China
[3]Department of Physics, University of California, Santa Barbara, CA, USA
[4]Department of Mechanical Engineering, Columbia University, New York, NY, USA
[5]National Institute for Materials Science, Tsukuba, Japan
[6]Laboratory of Atomic and Solid State Physics, Cornell University, Ithaca, NY, USA
[7]Kavli Institute at Cornell for Nanoscale Science, Ithaca, NY, USA

**Two-dimensional semiconductor moiré materials have emerged as a highly controllable platform to simulate and explore quantum condensed matter [1-19]. Compared to real solids, electrons in semiconductor moiré materials are less strongly attracted to the moiré lattice sites [1,2], making the nonlocal contributions to the magnetic interactions as important as the Anderson super-exchange [20]. It provides a unique platform to study the effects of competing magnetic interactions [20,21]. Here, we report the observation of strongly frustrated magnetic interactions in a Wigner-Mott insulating state at 2/3 filling of the moiré lattice in angle-aligned WSe$_2$/WS$_2$ heterobilayers. Magneto-optical measurements show that the net exchange interaction is antiferromagnetic for filling factors below 1 with a strong suppression at 2/3 filling. The suppression is lifted upon screening of the long-range Coulomb interactions and melting of the Wigner-Mott insulator by a nearby metallic gate. The results can be qualitatively captured by a honeycomb-lattice spin model with an antiferromagnetic nearest-neighbor coupling and a ferromagnetic second-neighbor coupling. Our study establishes semiconductor moiré materials as a model system for the lattice-spin physics and frustrated magnetism [22].**

Moiré materials formed by semiconducting transition metal dichalcogenides (TMDs) provide a physical realization of the extended Hubbard models [1, 3, 4, 6-11, 13-16, 23]. Correlated insulating states, including both Mott insulators [3-6, 11, 15, 16] and Wigner-Mott insulators [4, 11, 13, 24], have been recently observed in the strong correlation limit. Because the charge degree of freedom is frozen, the ground states and low-energy excited states of these correlated insulators are defined by the magnetic exchange interactions between localized spins [1, 7, 9]. Compared to real solids, electrons in the moiré lattice experience much weaker periodic potentials that are well approximated by harmonic potentials, and the Wannier functions extend over an appreciable portion of a unit cell [1, 2, 7, 10]. This makes the nonlocal contributions to the exchange interactions that are usually negligible in atomic-scale Hubbard systems, relevant. Recent theoretical studies show that a ferromagnetic (FM) direct exchange contribution (that is proportional to interaction) can be tuned to compete with an antiferromagnetic (AF) super-exchange contribution (that is inversely proportional to interaction) to induce a rich phase diagram [20, 21]. But the impact of nonlocal interactions on the moiré spin-physics has not been examined experimentally.

Here, we focus on the Wigner-Mott insulator state at 2/3 filling of the moiré lattice to reveal the frustrated magnetic interactions in angle-aligned $WSe_2/WS_2$ heterobilayers by magneto-optical measurements (Fig. 1). Angle-aligned $WSe_2/WS_2$ heterobilayers form a triangular moiré lattice with a period of about 8 nm [3, 4, 25]. The heterobilayers have a type II band alignment (Fig. 1b). The physics of the topmost moiré valence band can be mapped to a single-band triangular-lattice Hubbard model with a locked spin-valley pseudospin [1] (referred to simply as spin below). The size of the electron Wannier function is about 2 nm [1, 7]. The system is in the strong correlation limit with the nearest-neighbor hopping constant ($t \approx 1 - 2$ meV) much smaller than both the on-site ($U \approx 100 - 200$ meV) and the nearest-neighbor Coulomb repulsion ($V \approx 50$ meV) [1, 11].

We study two types of devices with a top and bottom gate (Fig. 1a). In both types, the gates are made of either graphite or $TaSe_2$ (a TMD metal) electrodes and hexagonal boron nitride (hBN) dielectrics. The gates can continuously tune the hole density in the moiré heterobilayer; filling factor $v = 1$ denotes one hole per moiré site or half-filled moiré valence band. In device 1, both gate electrodes are about 20 nm away from the heterobilayer and have a negligible screening effect. In device 2, the bottom gate electrode ($TaSe_2$) is placed about 1 nm below the heterobilayer, which practically screens out all but the on-site Coulomb repulsion in the moiré lattice. We identify the magnetic response associated with the Wigner-Mott insulating states by contrasting the behavior of the two types of devices. Details on the device fabrication and measurements are provided in Methods.

Figure 1c shows the filling-dependent reflectance contrast spectrum of device 1 near the fundamental moiré exciton resonance [26, 27] of $WSe_2$ at temperature $T = 1.7$ K. An enhancement in the exciton spectral weight can be identified at $v = 1$, 2/3 and 1/3. It arises, respectively, from the formation of a Mott or charge-transfer insulator [7] at $v = 1$ (Fig. 1d) and Wigner-Mott insulators (or generalized Wigner crystals [4]) at $v = 2/3$ (Fig. 1e) and 1/3. The latter is a manifestation of the strong long-range Coulomb repulsion $V$, whereas the Mott state is induced by the strong on-site Coulomb repulsion $U$. The two Wigner-Mott insulators melt around 35 K from temperature dependence studies. These results are consistent with the previous reports [4, 11, 13].

We examine the magnetic response of the moiré heterobilayers by performing the magnetic circular dichroism (MCD) measurements near the fundamental moiré exciton resonance of $WSe_2$ under an out-of-plane magnetic field (Fig. 2). The MCD is the difference between reflection of left and right circularly polarized incident light normalized by the total reflection. It provides a measure of the magnetization, $M$, of the moiré heterobilayer because of the valley-dependent optical selection rules and spin-valley locking in monolayer TMDs [3, 28, 29]. The magnetic field dependent MCD spectrum in Fig. 2a is from device 1 at $v = 2/3$ and $T = 1.7$ K. The signal is enhanced near the fundamental moiré exciton resonance (~ 1.67 eV). We use the integrated MCD over a narrow spectral window around the resonance (given by the two vertical dashed lines) to represent $M$; the analysis has been employed by earlier studies [3, 15].

Figure 2b shows the magnetic-field dependence of the integrated MCD of $\nu = 2/3$ at representative temperatures. The MCD increases linearly with magnetic field $B$ for small fields and saturates around 1 T at 1.7 K. As temperature increases, the MCD decreases and no clear saturation is observed within the field range of 2 T at high temperatures. We extract the magnetic susceptibility, $\chi \equiv \lim_{B \to 0} \frac{M}{B}$, from the MCD slope at zero field. Its temperature dependence is summarized in Fig. 2d. The susceptibility follows the Curie-Weiss (CW) law (solid line), $\frac{1}{\chi} = \frac{T-\theta}{C}$, where the Curie constant $C$ is proportional to the saturation magnetization and $\theta$ ($\approx$ -1.4 K) is the CW temperature. The CW law is known to describe the high-temperature magnetic response of interacting local moments with the magnitude $|\theta|$ reflecting the net exchange interaction energy $J$ between the local moments [30, 31] (we adopt the convention of expressing $J$ in kelvin). A negative $\theta$ corresponds to an AF exchange interaction. We limit the CW analysis to 5 – 35 K (above $|\theta|$ but below the melting temperature of the Wigner-Mott insulator), where the picture of local moments applies. The fitting results are, however, not sensitive to the precise choice of the temperature range. The picture of local moments is also consistent with Fig. 2b, which reflects the alignment and saturation of the local moments under an external magnetic field. At low temperatures, the saturation field ($B_s \approx 1$ T) also reflects $J$: a higher Zeeman field is required to overcome a stronger AF exchange to achieve magnetic saturation. At temperatures above $J$, the magnetic response is strongly affected by thermal excitations.

We perform similar measurements and analysis for the entire filling range of $\nu \leq 1$ (Extended Data Fig. 1 and 2). Figure 2c shows the filling-dependent magnetic susceptibility at varying temperatures. At high temperatures, a nearly linear dependence is observed (Extended Data Fig. 6). This reflects that the density of local moments increases with filling, $C \propto \nu$, as expected. As temperature decreases below $J$, $\chi$ increases drastically and a non-monotonic filling dependence emerges (Fig. 2c and 3a). Concurrently, both the CW temperature (Fig. 3b) and the saturation field at 1.7 K (Fig. 3c) also exhibit a strongly non-monotonic filling dependence. For all $\nu$'s, the CW temperature is negative or close to 0. The three quantities, $\chi$, $\theta$ and $B_s$, are fully consistent with each other. For instance, a small $|\theta|$ is accompanied by a small $B_s$ and a large $\chi$; they are manifestations of a small AF exchange interaction $J$.

The most striking feature of Fig. 3 is the strong suppression of the AF exchange around $\nu = 2/3$. The width of the feature in filling factor agrees well with the width of the Wigner-Mott insulator state at $\nu = 2/3$ (Fig. 1c). Similar effect is also likely around $\nu = 1/3$, but because the local moments are far apart, the exchange energy scale is too small to be fully resolved in this experiment. Suppression of the AF exchange at other fractional fillings between 1/3 and 2/3 is not observed. Because magnetism at generic fillings involving itinerant carriers is too complex a problem to be resolved in one study, we focus on fillings around 2/3 below.

To elucidate the role of the Wigner-Mott insulator on the magnetic response at 2/3 filling, we compare the behavior of device 1 and device 2. In device 2 (Fig. 1a), the bottom gate electrode effectively screens out the long-range Coulomb repulsion and quenches the Wigner-Mott insulator state [32]. This is consistent with recent reports [33, 34] and is supported by the absence of any enhancement in the exciton reflectance contrast at $v$ = 2/3 (Extended Data Fig. 3). No longer observable in device 2 are also the pronounced peak in $\chi$ and the dip in $B_s$ around 2/3 filling at 1.7 K (Fig. 3a, c). In addition, we are able to perform the CW analysis on the temperature dependence of the magnetic susceptibility (Extended Data Fig. 4). The presence of strong correlation likely contributes to the applicability of the CW analysis for interacting local moments here [35, 36]. The extracted CW temperature (Fig. 3b) is again negative with larger magnitude for the entire filling range ($\theta \approx$ -7 K at $v$ = 2/3). The dip in the filling dependence of $|\theta|$ around $v$ = 2/3 is no longer observable. All these results indicate that the observed suppression of the AF exchange interaction at 2/3 filling in device 1 is associated with the Wigner-Mott insulating state.

We consider the exchange mechanisms between local moments. The Anderson super-exchange, which involves virtual hopping of particles between two neighboring sites, is AF. When the long-range interactions are negligible, the super-exchange is inversely proportional to the on-site Coulomb repulsion ($\sim \frac{t^2}{U}$). This mechanism contributes to the observed net AF exchange and its enhancement upon screening of the Coulomb repulsions. But the super-exchange mechanism alone cannot explain the observed minimum in the filling dependence of the net AF exchange around 2/3 filling (Fig. 3b). Other contributions, such as the interaction-assisted hopping and direct exchange, have to be considered [20, 21]. These nonlocal contributions to the exchange interaction are relevant in semiconductor moiré materials because the harmonic trapping potentials are relatively shallow and the electron Wannier functions are extended [1, 2, 7, 10]. In particular, the direct exchange is FM and proportional to Coulomb interactions [20]; it can be tuned by screening to compete with the AF super-exchange to induce frustrated magnetism.

In the Wigner-Mott insulating state at $v$ = 2/3, holes form a crystal with a honeycomb lattice [24], which is 2/3 of the lattice sites of the original triangular moiré lattice (Fig. 1e). Because the system remains insulating when immediately doped away from the 2/3 filling (shown by low-temperature compressibility measurements of similar moiré heterobilayers [37]), the spin physics near the 2/3 filling can be captured by a (doped) spin model on the honeycomb lattice (Methods). The spin model is a good approximation in the flat band limit, as in angle-aligned WSe$_2$/WS$_2$ (Ref. [3, 4]). For simplicity, we consider a model that has an AF nearest-neighbor coupling $J_1$ and a FM second-neighbor coupling $J_2$ (both $J_1$ and $J_2$ are taken to be positive). The choice of a FM $J_2$ is consistent with the observed increase in $|\theta|$ when the long-range interaction is screened (Fig. 3b). Exactly at $v$ = 2/3, there are three nearest neighbors and six second neighbors per hole. The CW temperature, which is a weighted sum of the exchange interactions [30, 31], is hence given by $\theta = -(J_1 - 2J_2)$. The exchange is always AF as long as $J_1 > 2J_2$.

For fillings slightly above 2/3, the extra holes naturally reside at the center moiré sites of the honeycomb (Fig. 1f). They increase the AF nearest-neighbor links, but do not increase FM second-neighbor links. This simple picture implies that filling above 2/3 drives the system 'more AF'. Particularly, at filling factor 2(1+$p$)/3 with $p \ll 1$, the CW temperature can be expressed as $\theta = -[(J_1 - 2J_2) + (3J_1 + 2J_2)p]$ to the leading order in $p$. Its amplitude $|\theta|$ increases with $p$, which is consistent with the experimental data.

For fillings slightly below 2/3, vacancies are introduced into the honeycomb lattice (Fig. 1g). Introducing a vacancy removes twice as many FM links as AF links. This is identical to the original lattice and hence cannot explain the stronger AF exchange observed in experiment. However, if we further assume that the vacancies always reside on a pair of nearest-neighbor sites (Fig. 1g), introducing a pair of vacancies removes twelve FM links but only five AF links, and hence drives the system more AF. In particular, at filling factor 2(1-$p$)/3, the CW temperature can be expressed as $\theta = -[(J_1 - 2J_2) - \left(\frac{2J_1}{3} - 2J_2\right)p]$. As long as $3J_2 > J_1 > 2J_2$, $\theta$ is always negative and $|\theta|$ increases with $p$. We note that we do not attempt to prove that vacancies are indeed introduced in pairs in WSe$_2$/WS$_2$ moiré heterobilayers but a recent theoretical study on TMD moiré materials has suggested that doping in pairs could cost less Coulomb repulsion energy under the right condition [8]. Nevertheless, the spin model with only $J_1$ and $J_2$ adopted here is likely too simple to account for all the experimental data near the 2/3 filling.

In conclusion, we reveal the frustrated magnetic interactions in semiconductor moiré heterobilayers in the strong interaction limit through the highly non-monotonic filling dependence of the CW temperature. The strongly suppressed AF exchange in the Wigner-Mott insulator state at $\nu = 2/3$ can be qualitatively captured by a simple spin model that includes a competing AF nearest-neighbor and FM second-neighbor coupling. We do not observe any clear enhancement or suppression of the net exchange interaction at other commensurate fractional fillings between 1/3 and 2/3 where charge-ordered states have been reported [11, 13, 14]. Future studies are required to better understand the mechanism of magnetic exchange in the presence of charge order, nematicity [14] (stripe phases reported for the entire filling range of 1/3 < $\nu$ < 2/3) and itinerant electrons.

**Methods**
**Sample and device fabrication**
Dual-gate devices of angle-aligned WSe$_2$/WS$_2$ heterobilayers were fabricated using the dry-transfer method reported in the literature [3]. In short, atomically thin flakes of each constituent were exfoliated from bulk crystals onto SiO$_2$/Si substrates, and identified by their reflectance contrast under an optical microscope. The thickness was determined more accurately by the atomic force microscopy (AFM). A polymer stamp was used to pick up all of the flakes sequentially. The finished stack was released onto SiO$_2$/Si substrates with pre-patterned Au electrodes. Second harmonic generation (SHG) was employed to determine the crystal axes of WSe$_2$ and WS$_2$ monolayers before stacking [3]. For the second type of devices, an extra few-layer TaSe$_2$ flake was introduced into the device (Fig. 1a). It is separated from the WSe$_2$/WS$_2$ heterobilayer by a bilayer hBN spacer. The TaSe$_2$ flakes were exfoliated inside a nitrogen-filled glovebox to avoid sample degradation.

**Magneto-optical measurements**

Details of the magneto circular dichroism (MCD) measurements have been reported in Ref. [3, 15]. In short, the devices were mounted in a closed-cycle cryostat with a superconducting magnet (attoDRY 2100). The magnetic field was applied perpendicular to the sample plane. White light from a tungsten halogen lamp was collimated and focused onto the devices (with power less than 1 nW). The reflected light was detected by a liquid nitrogen-cooled charge-coupled device (CCD) attached to a grating spectrometer. The polarization of the white light was controlled by a combination of a polarizer and a broadband quarter-wave plate. The WSe$_2$/WS$_2$ heterobilayers were grounded during the optical measurements. In device 1, both the top and bottom gate voltages were controlled independently by two Keithley sourcemeters. In device 2, the TaSe$_2$ flake was grounded, and only the top gate voltage was applied to tune the filling factor.

The MCD spectrum is defined as $\text{MCD} = \frac{I_+ - I_-}{I_+ + I_-}$, where $I_+$ and $I_-$ are, respectively, the spectrum of the reflected left- and right-handed incident light from the sample. The integrated MCD signal is computed as $\frac{\int_{\epsilon_1}^{\epsilon_2} d\epsilon (\text{MCD})}{\int_{\epsilon_1}^{\epsilon_2} d\epsilon}$, where $\epsilon$ is the photon energy and the limits of integration are given by the dashed lines in Fig. 2a. We focus on the MCD response in this study rather than the exciton Zeeman splitting as in an earlier study [3]. The MCD directly reflects the optical Hall response of the system, and is straightforward to interpret [15]. In particular, the MCD is easier to analyze than the exciton Zeeman splitting in the presence of multiple exciton peaks at $\nu = 2/3$ which partially overlap in energy.

**Determination of the saturation magnetic field**

We use the Brillouin function, $\frac{M}{M_S} = \tanh(B/B_S)$, to describe the magnetic-field dependence of the integrated MCD signal or the sample magnetization $M$. Here $B$ denotes the applied magnetic field, respectively, $M_S$ and $B_S$ are the saturation magnetization and magnetic field, respectively. An example is shown in Extended Data Fig. 5 for $\nu = 0.96$ and T = 1.7 K. The filling dependence of $B_S$ is shown in Fig. 3c.

**Spin model**

At 2/3 filling, the charges form a Wigner-Mott crystal with the shape of honeycomb lattice [24], which is 2/3 of the lattice sites of the original triangular moiré lattice (Fig. 1e). We consider filling factors around 2/3. Although in our experiment the moiré lattice is filled with holes for the entire doping range, we refer to fillings slightly below and above 2/3 as, 'electron doping' and 'hole doping', respectively. One key observation we make is that, when immediately doped away from 2/3 filling, the system remains insulating at low temperature based on capacitance measurements [37]. This implies that the doped charges in the flat band system are always immobile, likely due to the presence of disorder and/or strong correlation. The spin physics of the system near 2/3 filling is therefore captured by a (doped) spin model on the honeycomb lattice with each hole carrying a spin-1/2 degree of freedom.

The spin model we adopt has antiferromagnetic (AF) nearest-neighbor (nn) coupling $J_1$, and ferromagnetic (FM) next-nearest-neighbor (nnn) coupling $J_2$ (in our convention both $J_1$ and $J_2$ are taken to be positive). The Hamiltonian of the spin model takes the following form:

$$H = \sum_{nn\,bonds} J_1 \vec{S}_i \cdot \vec{S}_j - \sum_{nnn\,bonds} J_2 \vec{S}_i \cdot \vec{S}_j + \sum_{sites} \vec{h} \cdot \vec{S}_i, \tag{1}$$

where $\vec{S}_i$ is the spin on site $i$ and $\vec{h}$ is the external magnetic field. Since the main physics we focus on is observed at finite temperature, we will use the high temperature expansion to analyze this model. Note that the qualitative results of our high-temperature expansion calculation are insensitive to whether the spin is treated as a classical three-component vector or a quantum spin-1/2 variable. The following calculation is performed using the former treatment.

With hole doping away from 2/3 filling, namely $\nu > 2/3$, there are two main qualitative findings in experiments: (1) The Curie-Weiss (CW) temperature $\theta$ remains negative, but its magnitude increases under hole doping (Fig. 3b); this implies that overall speaking the system is AF, and it becomes even 'more AF' under hole doping. (2) The spin susceptibility decreases with hole doping at low temperature (Fig. 3a), but increases with hole doping at high temperature (Extended Data Fig. 6).

Figure 1f depicts the setup for hole doping away from 2/3 filling. The extra hole doped to the system naturally reside at the center moiré sites of the honeycomb. A doped hole increases the nearest-neighbor AF links in the system (labelled as dashed lines), but does not increase FM next-nearest-neighbor links. This simple picture implies that hole doping drives the system 'more AF', and increases $|\theta|$. At filling $\nu = 2(1+p)/3$, the high temperature expansion leads to the following result of $\theta$, expanded to the leading order of doping $p$ ($> 0$):

$$\theta = -[(J_1 - 2J_2) + (3J_1 + 2J_2)p] + O(p^2). \tag{2}$$

As long as $J_1 > 2J_2$, $\theta$ is always negative, signifying AF behavior. Also, $|\theta|$ increases with hole doping $p$ at the leading order expansion of $p$, which is consistent with the experimental data. The spin susceptibility $\chi$ reads

$$\chi = \frac{n(1+p)}{3(T-\theta)} \sim \frac{n(1+p)}{3T} + \frac{n(-J_1 + 2J_2 - 4J_1 p)}{3T^2} + O(p^2), \tag{3}$$

where $n$ is the number of holes of the honeycomb at $\nu = 2/3$.

As an example, let us choose $J_1 = 2.5 J_2 = 2.5A$, we then have $\theta \approx -0.5A$. One immediately notices that when $T < 10A$, $\chi$ decreases with doping $p$; while $T > 10A$, $\chi$ increases with $p$. All these are consistent with the experimental observations.

With electron doping away from 2/3 filling, namely $\nu < 2/3$, there are also two main qualitative findings in experiments: (1) The CW temperature $\theta$ remains negative; and the magnitude of $|\theta|$ increases under hole doping (Fig. 3b); this implies that the system still becomes 'more AF' under hole doping. (2) The spin susceptibility always decreases with doping within the temperature range in experiment (Fig. 3a).

Electron doping will be modeled as doping vacancies in our spin model. The experimental facts are somewhat counter-intuitive, as doping vacancies normally does not increase $|\theta|$. However, in the following we will show that with a simple extra assumption of the configurations of doped vacancies, our model can explain the two experimental facts listed above. The setup for vacancies doping is depicted in Fig. 1g: we assume that the doped vacancies always reside on a pair of nearest-neighbor sites. Although we do not prove here that the vacancies do form pairs, we note that a similar pair-wise doping in TMD moiré materials in the flat band limit has been discussed by Ref. [8] as a way to minimize the Coulomb repulsion energy. At filling $\nu = 2(1-p)/3$, the high temperature expansion calculation gives the following results

$$\theta = -[(J_1 - 2J_2) - \left(\frac{2J_1}{3} - 2J_2\right)p] + O(p^2). \qquad (4)$$

As long as $3J_2 > J_1 > 2J_2$, $\theta$ is always negative, and $|\theta|$ increases with $p$, again consistent with experimental data. The reason $|\theta|$ increases with $p$ is that, on the original lattice model, there are twice as many second neighbor links (FM interactions) as nearest-neighbor links (AF interactions). Since the vacancies are 'paired up', doping a pair of vacancies remove 12 FM links, but remove only 5 AF links. Hence doping vacancies can indeed drive the system more AF, and increase $|\theta|$. The spin susceptibility $\chi$ is calculated as

$$\chi = \frac{n(1-p)}{3(T-\theta)} \sim \frac{n(1-p)}{3T} + \frac{n[-J_1 + 2J_2 + \left(\frac{5J_1}{3} - 4J_2\right)p]}{3T^2} + O(p^2), \qquad (5)$$

Again, as an example let us choose $J_1 = 2.5J_2 = 2.5A$. Then we obtain $\theta \approx -0.5A$. We found that as long as $T > A/6$, $\chi$ always decreases with $p$, which is consistent with experimental findings.

**Acknowledgements**
We thank Veit Elser, Liang Fu, Eun-Ah Kim and Steven Kivelson for helpful discussions.

## Figures

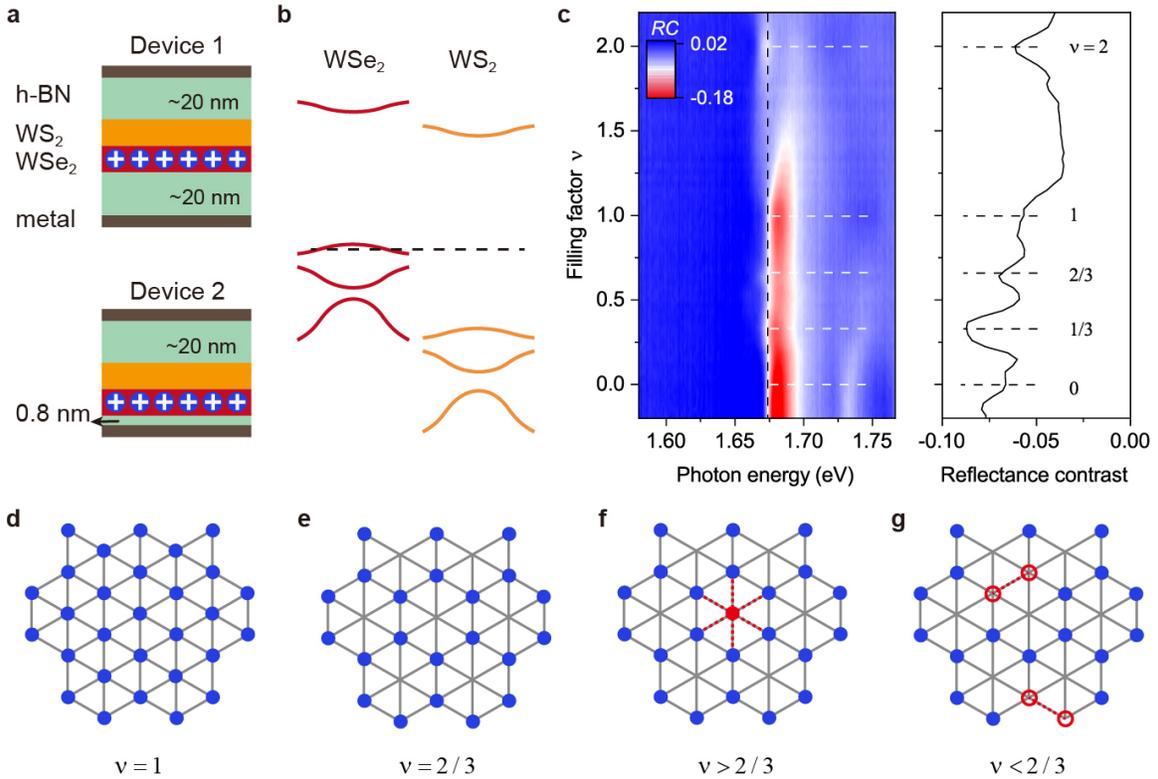

**Figure 1 | Correlated insulating states in WSe$_2$/WS$_2$ moiré lattices. a,** Cross-sectional schematics of two types of devices investigated in this study. The WSe$_2$/WS$_2$ moiré lattice is hole-doped with the holes (positive charges) residing in the WSe$_2$ layer. The bottom gate electrode is about 20 nm and 1 nm below the WSe$_2$ layer in device 1 and 2, respectively. **b,** Schematic moiré band structure. The Fermi level (dashed line) is inside the first moiré valence band of WSe$_2$. **c,** Filling-dependent reflectance contrast (RC) spectrum of device 1 at 1.7 K (left). RC enhancement near the fundamental moiré exciton resonance of WSe$_2$ is observed at several commensurate filling factors (dashed horizontal lines). The right panel shows a linecut at 1.674 eV. **d-g,** Charge configurations at zero temperature on the underlying triangular moiré lattice for $\nu = 1$ (**d**), 2/3 (**e**), slightly above 2/3 (**f**) and below 2/3 (**g**). Filled circles denote occupied sites. In **f**, an extra hole resides at the center site of a honeycomb formed by the occupied sites; in **g**, the doped vacancies (empty circles) are assumed to reside on a pair of nearest-neighbor sites in a spin model to account for the magnetic susceptibility near $\nu = 2/3$.

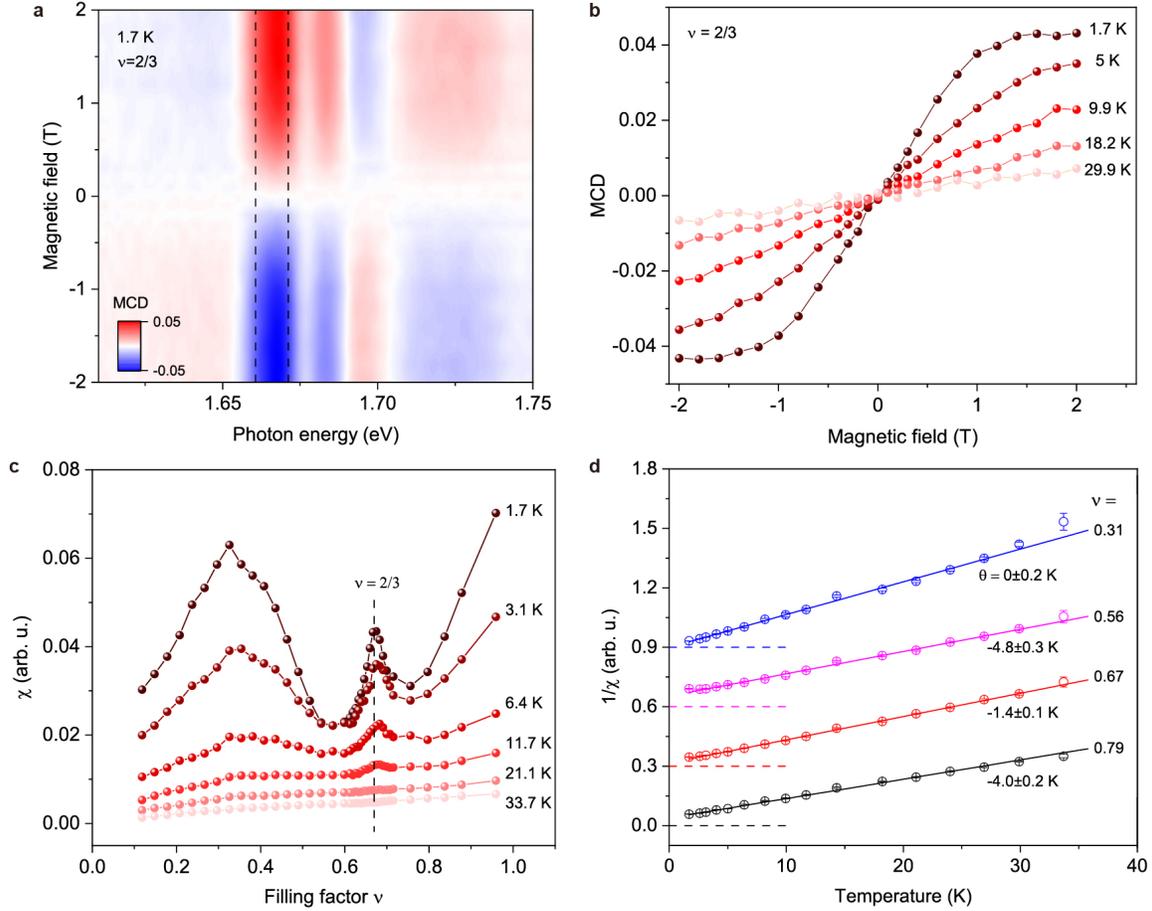

**Figure 2 | Magneto optical measurements (device 1). a,** MCD spectrum as a function of magnetic field at $\nu = 2/3$ and temperature $T = 1.7$ K. The spectrum is integrated over a narrow window around the fundamental moiré exciton resonance (within the two vertical lines) to represent the sample magnetization. **b,** Magnetic-field dependence of integrated MCD at representative temperatures. The MCD saturates above ~ 1 T at 1.7 K. The slope at zero magnetic field is extracted to represent the magnetic susceptibility. **c,** Filling dependence of the magnetic susceptibility, $\chi$, at varying temperatures. A sharp peak emerges at 2/3 filling as temperature decreases. **d**, Curie-Weiss analysis of the temperature dependence of the inverse susceptibility (symbols). The error bars are propagated from the uncertainties of the linear slope at zero field in **b**. The CW temperatures corresponding to the best fits (solid lines) are included. Data for different filling factors are vertically displaced for clarity. The horizontal dashed lines mark where $1/\chi$ is zero.

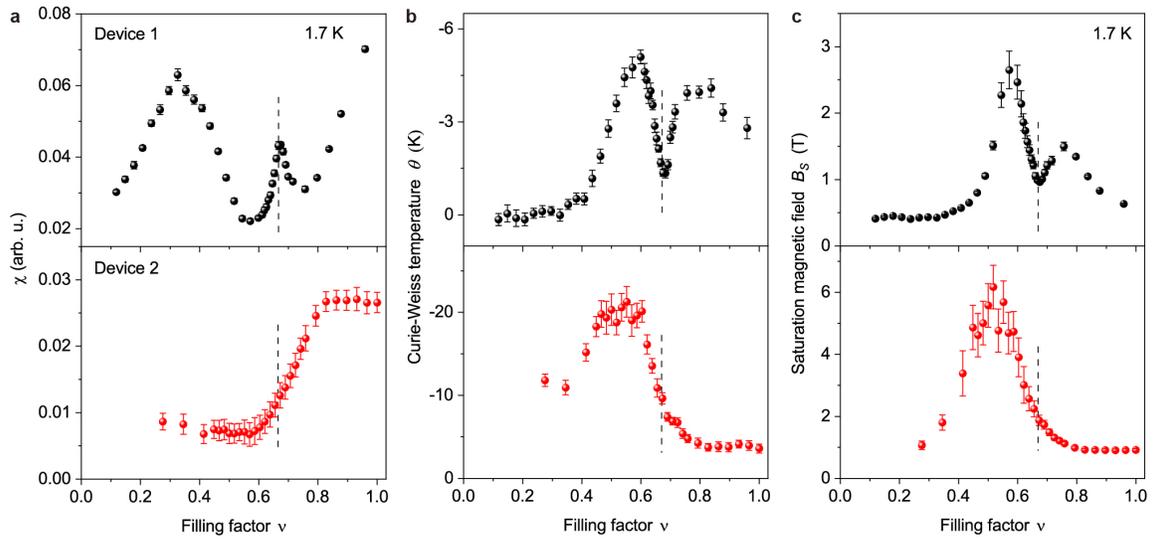

**Figure 3 | Comparison of the magnetic response of device 1 and device 2. a-c**, Filling dependence of the magnetic susceptibility at 1.7 K (**a**), the CW temperature (**b**), and the saturation magnetic field at 1.7 K (**c**). Data without gate screening (device 1) and with screening (device 2) are shown in the top and bottom panels, respectively. The strongly suppressed AF exchange at 2/3 filling (dashed lines) is observed in device 1, but not in device 2. The error bars correspond to the uncertainties of the linear slope from the field dependence of the MCD at zero field (**a**), the CW analysis (**b**) and the Brillouin-function fit as described in Methods (**c**).

# Extended Data Figures

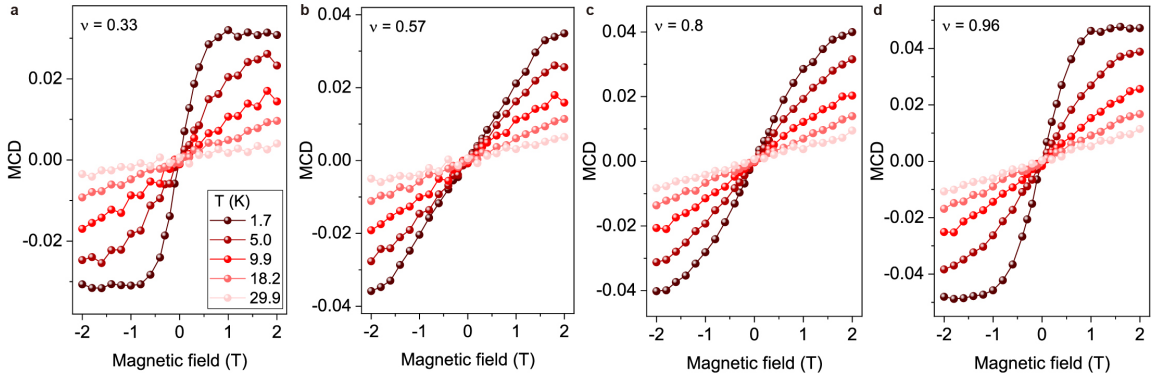

**Extended Data Figure 1 | Magnetic response of device 1 at varying temperatures for $\nu$ = 0.33 (a), 0.57 (b), 0.8 (c) and 0.96 (d).** The color of the symbols varying from darkest to lightest denotes the temperature from 1.7 to 29.9 K.

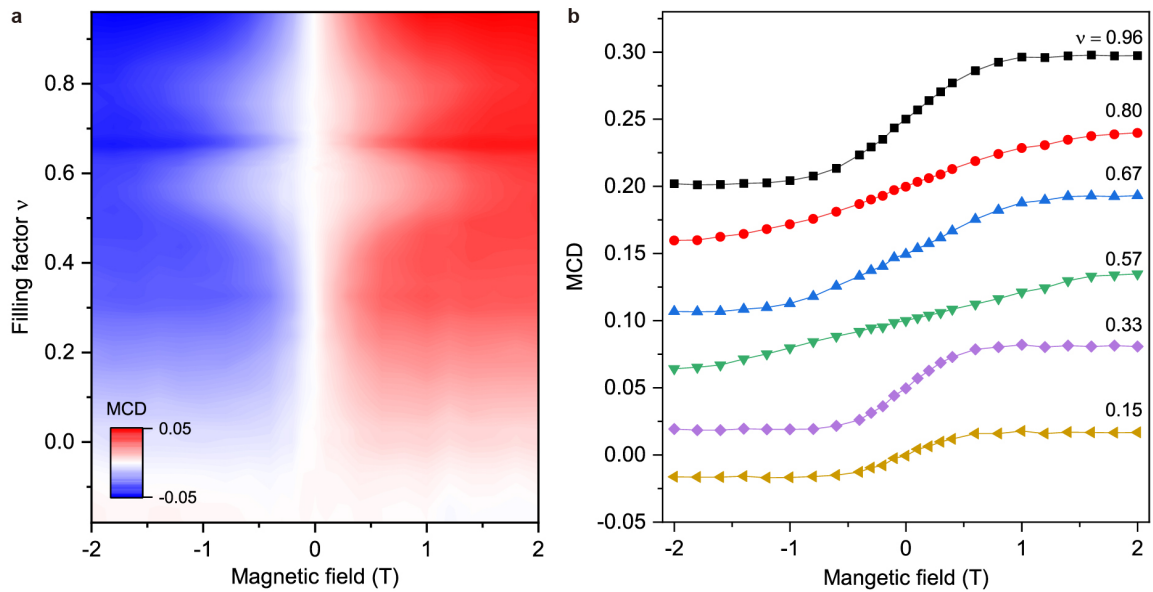

**Extended Data Figure 2 | Magnetic response of device 1 at 1.7 K. a,** Integrated MCD as a function of magnetic field and filling factor. **b,** Linecuts of **a** at selected filling factors. The curves are displaced vertically for clarity.

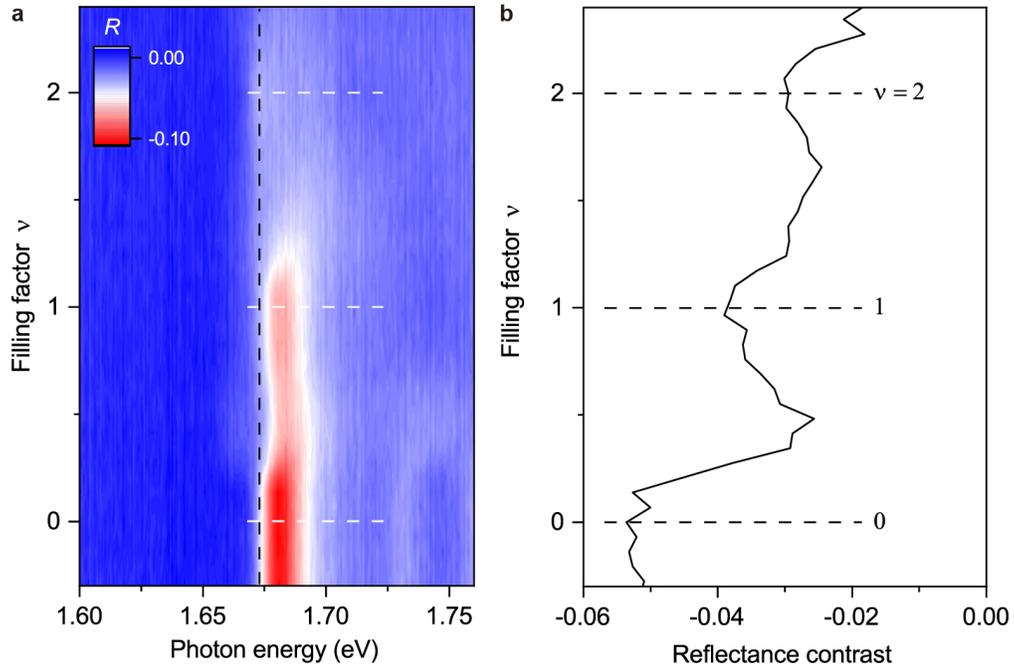

**Extended Data Figure 3 | Optical characterization of device 2 at 1.7 K. a,** Filling-dependent reflectance contrast (RC) spectrum. Enhanced RC near the fundamental moiré exciton resonance of $WSe_2$ is observed only at integer filling factors. The absence of the RC enhancement at $\nu = 1/3$ and $2/3$ supports the absence of these Wigner-Mott insulator states. **b,** Linecut of **a** at 1.673 eV.

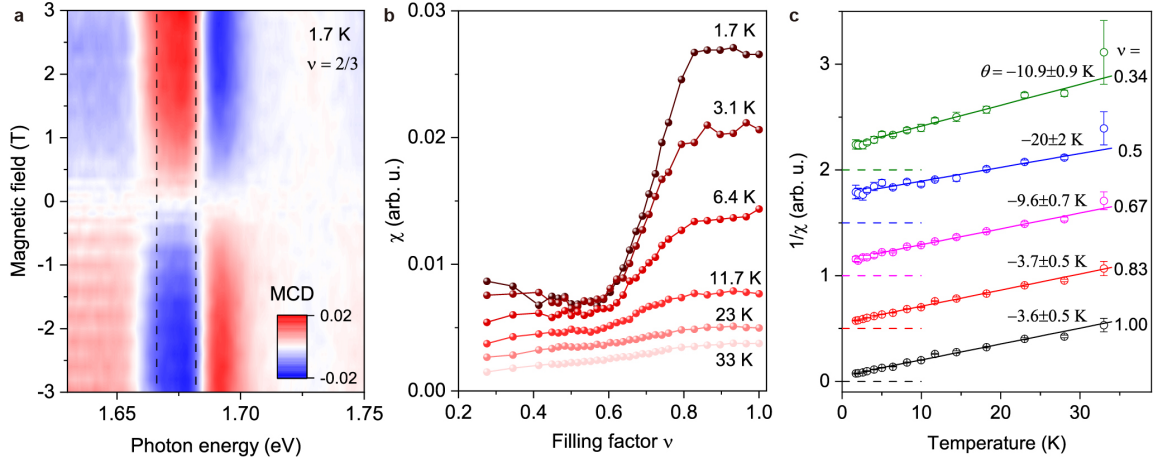

**Extended Data Figure 4 | Magneto-optical measurements (device 2). a,** MCD spectrum as a function of magnetic field at $\nu = 2/3$ and $T = 1.7$ K. The spectrum is integrated over a narrow window around the fundamental moiré exciton resonance (within the two vertical lines) to represent the sample magnetization. **b,** Filling dependence of the magnetic susceptibility, $\chi$, at varying temperatures. The susceptibility is extracted from the linear slope of the magnetic-field dependence of integrated MCD at zero field. The solid lines are guide to the eye. **c**, Curie-Weiss analysis (solid lines) of the inverse susceptibility (symbols) as a function of temperature. The error bars are propagated from the uncertainty of $\chi$. The CW temperatures corresponding to the best fits (solid lines) are included. Data for different filling factors are vertically displaced for clarity. The horizontal dashed lines mark where $1/\chi$ is zero.

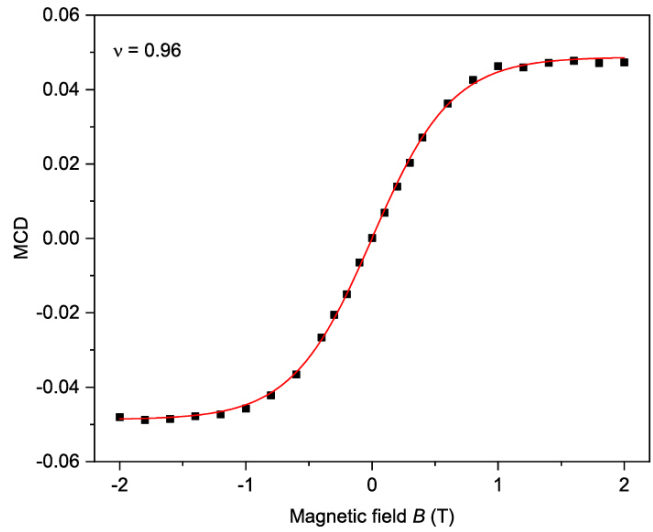

**Extended Data Figure 5 | Determination of the saturation magnetic field (device 1).** The red curve is a Brillouin-function fit to the integrated MCD as a function of magnetic field (at 0.96 filling and 1.7 K). The corresponding saturation magnetic field is $0.63 \pm 0.01$ T.

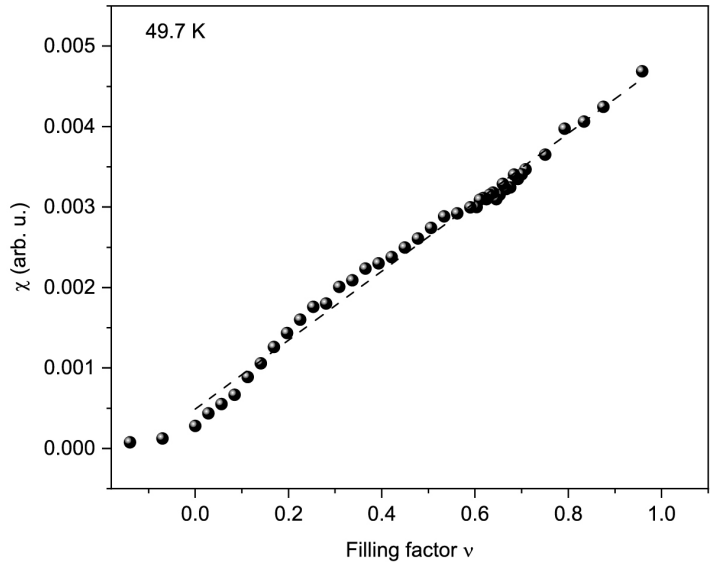

**Extended Data Figure 6 | The magnetic susceptibility as a function of fillings at high temperature (device 1).** The dashed straight line is just an eye guidance to the linear filling dependence.